\documentstyle[12pt]{article}
\begin{document}

\begin{center}
{\Large Hypertriton Electric Polarizability} \\[.4in]

\setlength{\baselineskip}{0.1in}
{\large V. F. Kharchenko}\footnote{E-mail: vkharchenko@bitp.kiev.ua}
{\large and A. V. Kharchenko} \\[.05in]

{\footnotesize \it Bogolyubov Institute for Theoretical Physics,\\
National Academy of Sciences of Ukraine, UA - 03143, Kyiv, Ukraine}\\[.4in]
\end{center}

\begin{abstract}
\setlength{\baselineskip}{0.1in}
\noindent
{\footnotesize A rigorous formalism for determining the electric dipole
polarizability of a three-hadron bound complex in the case that
the system has only one bound (ground) state has been elaborated.
On its basis, by applying a model wave function that takes into
account specific features of the structure of the lightest
hypernucleus and using the known low-energy experimental data for
the $p-n$ and $\Lambda-d$ systems as input data, we have calculated the
value of the electric dipole polarizability of the lambda hypertriton
$\alpha_E(^3_{\Lambda}\mbox{H})$. It follows from our study that
the polarizability of the lambda hypertriton is close to $3\;\mbox{fm}^3$
exceeding the polarizabilities of the ordinary three-nucleon nuclei
by an order of magnitude.}\\[.2in]
\end{abstract}

\vspace*{.1in}
\noindent {\bf 1. Introduction} \\ [.1in]
Data on the polarization (deformation) of the few-body
nuclear systems in the electromagnetic field that is a fundamental
property of each nucleus contain important information on the
nuclear force. To date, by studying deviations from Rutherford
scattering below the Coulomb barrier of a light nucleus in an
intense Coulomb field of a heavy nucleus, the values of the
electric dipole polarizability of the two lightest nuclei, of the
deuteron and the $^3\mbox{He}$ nucleus, have been directly
measured,
\begin{equation}
\alpha_E(^2\mbox{H})=0.70 \pm 0.05\;\; \mbox{fm}^3\;(\mbox{Ref.
1})\,,\;\;  \alpha_E(^3\mbox{He})=0.25 \pm 0.04\;\;
\mbox{fm}^3\;(\mbox{Ref. 2})\,.
\end{equation}

An other way  of obtaining the value $\alpha_E$, that is based on
the known relation between the polarizability and an energy-weighted
integral over the total photoabsorption cross-section for the
corresponding nucleus $\sigma$ (the sum rule $\sigma_{-2}$) with
the use of the experimental data for $\sigma$, leads to the values
$\alpha_E(^2\mbox{H})=0.61\pm 0.04\;\mbox{fm}^3\;\mbox{(Ref.
3)}\;\;\mbox{and}\;\;\alpha_E(^3\mbox{He})$ in the range from
$0.13\;\mbox{to}\;0.17\;\mbox{fm}^3\; \mbox{(Ref. 2)}\,$, supporting the
result of the direct measurement in the case of the deuteron but being
inconsistent with that in the case of the $^3\mbox{He}$ nucleus.

Calculations of the deuteron electric polarizability, carried out
in the framework of the traditional nuclear physics for the
realistic nucleon-nucleon interaction potentials, lead to the
values of $\alpha_E(^2\mbox{H})\;\mbox{in the range from}\;0.631\;\mbox
{to}\;0.634\;\mbox{fm}^3$ (Refs.  4 and 5), being consistent with experiment.
Furthermore, examining theoretically the anisotropy of the deformation of the
deuteron in the electric field caused by the tensor character of the $n-p$
interaction$^{5,6}$ each of the components of the deuteron electric
polarizability $\alpha^{|M|}_E$, the longitudinal component (with
the deuteron spin along the electric field) $\alpha^1_E$ and the
transverse one $\alpha^0_E$, have been individually
calculated,
\begin{equation}
\alpha^1_E(^2\mbox{H})=0.669\;\; \mbox{fm}^3\,,\;\;\mbox{and}\;\;
\alpha^0_E(^2\mbox{H})=0.555\;\; \mbox{fm}^3\,\; (\mbox{Ref. 5})\, .
\end{equation}
(The above electric polarizability $\alpha_E(^2\mbox{H})$ is the
averaged value of the components $\alpha^{|M|}_E$,
$\alpha_E(^2\mbox{H})=\frac{2}{3}\alpha^1_E +
\frac{1}{3}\alpha^0_E$ .) The similar components of the
paramagnetic dipole susceptibility of the deuteron have been
calculated in Ref. 7.

Computations of scalar and tensor deuteron polarizabilities have
also been performed in the framework of the effective field theory
that uses space-time and global chiral symmetries of the quantum
chromodynamics consistently describing pion propagation and
relativistic effects$^{8,9}$. Though the results for the electric
and magnetic deuteron polarizabilities obtained in the cited works
in the leading and next-to-leading orders agree with the values
calculated in the traditional nuclear physics with the application
of the potential models, it can look forward the occurence of
deviations from the values at higher orders in the effective field
theory expansion.

Presently, few-body methods are widely used in the physics of
hypernuclei --- hadronically stable bound formations of protons,
neutrons and one or more hyperons. Data on properties of
few-hadron systems are the basic source of obtaining information
on the interaction forces both between $\Lambda$ hyperon and
nucleon (${\Lambda}-N$ interaction) and between two
$\Lambda$ hyperons ($\Lambda-\Lambda$ interaction). To date,
several tens lambda hypernuclei have been found experimentally.
The very lightest hypernuclei are a three-hadronic system consisting
of the proton, the neutron and the $\Lambda$ hyperon in a bound state,
the hypertriton $^3_{\Lambda}\mbox{H}$, and the two mirror four-hadron
nuclei, $^4_{\Lambda}\mbox{H}$ and $^4_{\Lambda}\mbox{He}$, that are
bound $pnn\Lambda$ and $ppn\Lambda$ systems, respectively. The
hypertriton is the simplest halo nucleus, moreover, with the unique
strange halo. It can be considered as a weakly bound two-body
formation from the $\Lambda$ particle and a core in the form of the
deuteron which are separated by distances substantially exceeding
the deuteron size.

Of four double-$\Lambda$ hypernuclei reported thus far the
lightest are $^{\;\;\;4}_{\Lambda\Lambda}\mbox{H}$ (Ref. 10) and
$^{\;\;\;6}_{\Lambda\Lambda}\mbox{He}$ (Refs. 11 and 12). Evidence
on the production of the hypernucleus
$^{\;\;\;4}_{\Lambda\Lambda}\mbox{H}$ in (${K^{-},K^{+}}$) on
$^9\mbox{Be}$ has been presented after analysing the experiment
E906 completed at the Alternating Gradient Synchrotron of the
Brookhaven National Laboratory in Ref. 10. However, to date there
arises a doubt as to existence of the bound system
$^{\;\;\;4}_{\Lambda\Lambda}\mbox{H}$, since its formation and
decay were not needed to explain the data of this
experiment$^{13}$. Hence, the existence of hypernucleus
$^{\;\;\;4}_{\Lambda\Lambda}\mbox{H}$ has not been conclusively
proved yet.

This paper is devoted to study of the behaviour of the lightest of
all $\Lambda$ hypernuclei, the hypertriton $^3_{\Lambda}\mbox{H}$,
in the external Coulomb field. A new expression for the electric
dipole polarizability of the three-hadron bound complex, obtained
by us in the special case that the system has only one bound
(ground) state, is given in Section 2. The application of this
expression to the hypertriton in the framework of a physically
justified model is considered in Section 3. The results of
corresponding calculations of $\alpha_E(^3_{\Lambda}\mbox{H})$
based on the known low-energy data for the $p-n$ and $\Lambda-d$
interactions are discussed in Section 4.
Section 5 is devoted to a short summary and conclusions.\\

\vspace*{.1in}
\noindent {\bf 2. Electric polarizability of the three-hadron bound
system} \\ [.1in]
We derive an expression for the electric dipole
polarizability of a bound hadronic complex (that consists of $N$
particles), $\alpha_E$ , on the basis of the rigorous few-body
approach using the formalism of the effective interaction of the
charged complex with external Coulomb field, the source of which
can be any charged particle or nucleus (the particle 1). The
electric dipole polarizability of the complex characterizes the
strength of the effective potential of interaction between the
particle 1 and the centre of mass of particles of the complex at
asymptotically large distances. In the simplest event that the complex
contains only one charged particle (the particle 2) and the rest
are neutral, the electric dipole polarizability of the complex can
be written as
\begin{equation}
\alpha_E=-2<\Psi_0\mid ({\bf
D}_2\cdot\hat{\mbox{\boldmath$\rho$}}_1) G^Q(-B_0) ({\bf
D}_2\cdot\hat{\mbox{\boldmath$\rho$}}_1)\mid \Psi_0>\,,
\end{equation}
where $\Psi_0$, ${\bf D}_2=e_2{\bf
r}_2$, and $G^Q(-B_0)$ are the wave function of the ground bound
state of the complex (of $N$ particles) corresponding to the
binding energy $B_0$ (normalized to unit, $<\Psi_0\mid\Psi_0>=1$),
the operator of the dipole moment of the charged particle 2 (having
the charge $e_2$) of the complex, and the product of the projection
operator $Q=1-P$ ($P$ is the projection operator on to the complex
ground state, $P=\mid\Psi_0><\Psi_0\mid$) and the full Green's
operator of the complex $G(E)=(E-H_0-V)^{-1}$, $G^Q(E)=QG(E)$,
at the energy $E=-B_0$, respectively. Here $H_0$ is the kinetic
energy operator of the complex and $V$ is the total interaction
potential. The latter is given as $V=\sum_{i<j}^{} v_{ij}$
where $v_{ij}$ is the potential of the pair interaction between
the particles $i$ and $j$, the summation is over all the particles
of the complex $2,3,...N$.  The quantities ${\mbox{\boldmath$\rho$}}_1$
and ${\bf r}_2$ in Eq. (3) are the radius vectors specifying the
relative positions of the particle 1 forming the external Coulomb field
and the particle 2, the charged constituent of the complex, both with
respect to the centre of mass of the complex (the hat signifies a unit vector:
$\hat{{\mbox{\boldmath $\rho$}}}_1\equiv{\mbox{\boldmath $\rho$}}_1/\rho_1$).

Previously, on the basis of the three-body formalism of the
effective interaction of a charged particle and a bound
complex$^{14-17}$, we have derived an expression for the
polarization potential of the two-hadron ($N=2$) bound complex
that consists of charged and neutral hadrons (for example,
the deuteron)$^{15,18,19}$, starting immediately from the Faddeev
integral equations$^{20}$. In the case that the interaction
between the proton (the particle 2) and the neutron composed the
deuteron is central, the general formula (3) for the electric
dipole polarizability of the deuteron is reduced to
\begin{equation}
\alpha_E(^2\mbox{H})=\frac{2}{3}\frac{e^2}{\hbar^2 c^2}
\left(\frac{m_n}{m_{pn}}\right)^2
\int_{0}^{\infty} \frac{dk k^2}{2\pi^2} \frac{\mid \psi_0^\prime (k)\mid^2}
{\frac{k^2}{2\mu_{pn}}+B_d},
\end{equation}
where $e$ is the charge of the proton, $\mu_{pn}$ is the
proton-neutron reduced mass, $\mu_{pn}=m_pm_n/m_{pn}$,
$m_{pn}=m_p+m_n$ ($m_p$ and $m_n$ are the proton and neutron
masses), $\psi_0^\prime(k)\equiv d\psi_0(k)/dk$ is the first
derivative of the deuteron wave function in the momentum space in
the variable of magnitude of the relative momentum $k$ and $B_d$
is the binding energy of the deuteron. The formula (4) is in
agreement with the expression for $\alpha_E$ obtained in the case
of the separable S-wave pair potential in Refs. 15, 18 and 19.

In this work, to derive the formula for the electric dipole
polarizability of the hypertriton $^3_{\Lambda}\mbox{H}$ ($N = 3$),
we start from the four-body problem, considering the bound system
of the charged particle --- the proton (the particle 2) --- and two
neutral particles --- the neutron (the particle 3) and the
$\Lambda$ hyperon (the particle 4) --- in the Coulomb field of the
charged particle 1. Taking into consideration the analytical
properties of the three-body transition matrix that determines the
Green's function in Eq. (3), we find the following expression for
the electric dipole polarizability of the hypertriton:
\begin{equation}
\alpha_E(^3_{\Lambda}\mbox{H})=2\frac{e^2}{\hbar^2 c^2}
\int \frac{d{\bf k}d{\bf p}}{(2\pi)^6} \frac{\mid \left( \frac{m_n}{m_{pn}}
{\bf \nabla}_{\bf k}+
\frac{m_{\Lambda}}{m_{pn\Lambda}}
{\bf \nabla}_{\bf p}\right) \Psi_0 ({\bf k},{\bf p}) \mid ^2}
{\frac{k^2}{2\mu_{pn}}+\frac{p^2}{2\mu_{pn,\Lambda}}+B_{ht}},
\end{equation}
where $\mu_{pn,\Lambda}$ is the reduced mass of the proton-neutron
system with the mass $m_{pn}=m_p+m_n$ and the $\Lambda$ hyperon
with the mass $m_\Lambda$,
$\mu_{pn,\Lambda}=m_{pn}m_{\Lambda}/m_{pn\Lambda}$,
$m_{pn\Lambda}= m_p+m_n+m_{\Lambda}$, $B_{ht}=B_d+B_{\Lambda}$ is
the binding energy of the hypertriton that equals to the sum of
the deuteron binding energy $B_d=\kappa_d^2/{2\mu_{pn}}$ and the
binding energy of  the $\Lambda$ hyperon
$B_{\Lambda}=\kappa_{\Lambda}^2/{2\mu_{d,\Lambda}}$,
$\mu_{d,\Lambda}=m_d m_{\Lambda}/m_{d\Lambda}$, $m_d$ is the mass
of the deuteron, $m_{d\Lambda}=m_d+m_{\Lambda}$, $\Psi_0({\bf
k},{\bf p})$ is the normalized wave function of the hypertriton in
the momentum space, ${\bf k}$ and ${\bf p}$ being the Jacobi
momentum variables describing the relative motion of the proton
$p$ and the neutron $n$ and the motion of the centre of mass of
$p$ and $n$ relative to the particle $\Lambda$, respectively
\begin{equation}
{\bf k}=\frac{m_n {\bf k}_p-m_p {\bf k}_n}{m_{pn}},\;\; {\bf
p}=\frac{m_{\Lambda}({\bf k}_p+{\bf k}_n)-m_{pn}{\bf k}_{\Lambda}}
{m_{pn\Lambda}}\;,
\end{equation}
$k_i$ is the momentum of the particle $i$.

We emphasize that the formula for the electric polarizability of
the three-hadron system (5) has been derived on the assumption
that the interactions between constituents support the existence
of only one (ground) bound state of the system. As this takes place,
the formula (5) provides a rigorous treatment of the problem. It
is significant that the approach developed by us does not require
a knowledge of the three-body wave functions of the continuum states.\\

\vspace*{.1in}
\noindent {\bf 3. Wave function of the hypertriton} \\ [.1in]
Unlike the wave function of the triton, the wave
function of the hypertriton has a distinct cluster character. The
$\Lambda-N$ interaction force is weaker than the $n-p$ one to an
extent that each of the interaction potentials $v_{{\Lambda}p}$ and
$v_{{\Lambda}n}$ by itself does not form a bound state --- the
hyperdeuteron does not exist. Moreover, in the hypertriton the
$\Lambda$ hyperon is bound up with both nucleons much weaker than
the proton and the neutron are bound up together in the deuteron
($B_\Lambda/B_d=0.058$).The deuteron in the hypertriton is only
slightly deformed by the $\Lambda$ hyperon. Therefore, the hypertriton
may be considered as a two-cluster bound system composed of the
$\Lambda$ hyperon and a core in the form of the deuteron. Herewith,
the hypertriton size determined by the distance between the $\Lambda$
hyperon and the deuteron centre of mass substantially exceeds the
deuteron size. In accordance with the uncertainty principle, it follows
that in the hypertriton the motion of the $\Lambda$ hyperon relative
to the centre of mass of the deuteron proceeds slowly in comparison
to a quicker relative motion of the nucleons $p$ and $n$ inside
the deuteron:  $p/k\propto\kappa_\Lambda/\kappa_d=0.295$.

Under these conditions, it may be approximately considered that in
the hypertriton the relative proton-neutron motion and the
relative motion of the $\Lambda$ hyperon and the deuteron centre
of mass operates independently of one another. Then the wave
function of the hypertriton is factorized taking the form
\begin{equation}
\Psi_0({\bf k},{\bf p})=\psi_0(k) \phi_{\Lambda}(p),
\end{equation}
where $\psi_0(k)$ is the normalized wave function of the deuteron
in the ground state with the binding energy $B_d$ corresponding to
the pair interaction potential $v_{pn}$, and $\phi_{\Lambda}(p)$ is
the normalized wave function of the ground bound state of the $\Lambda$
hyperon and the deuteron with the binding energy $B_\Lambda$ that
corresponds to the effective $\Lambda-d$ interaction potential
$v_\Lambda$. The potential $v_\Lambda$ depends only on the radius-vector
of the particle $\Lambda$ relative to the centre of mass of the deuteron
\mbox{\boldmath $\rho$} (or on the variable momentum ${\bf p}$ in the
momentum space).

The expression (7)  can be considered as the first term of an
expansion of the three-particle function $\Psi_0$ in terms of the
complete set of wave functions of the complex  of the interacting
proton and neutron that corresponds to the ground bound state of
the complex. In Eq. (7) all the summands of the expansion in terms
of the functions of the continuous spectrum of the complex $p+n$
were neglected.

The dominance of the factorized term (7) in the hypertriton wave
function has been substantiated rigorously by the three-body
calculations with the use of the realistic potentials in Ref. 21.
According to them, the integral of the square of the overlap of
the deuteron with the hypertriton normalized wave functions
$\phi_\Lambda(p)\equiv<\psi_0\mid\Psi_0>$ taken over all values
of the variable that describes the relative motion of $\Lambda$
and the centre of mass of the deuteron,
$P_d(^3_{\Lambda}\mbox{H})\equiv<\phi_\Lambda\mid\phi_\Lambda>$, is found
to be $P_d(^3_{\Lambda}\mbox{H})=0.987$. From this result, close to unity
ensured by the model (7), the cluster character of the hypertriton
wave function is immediately evident. For comparison, in the case of
the triton the integral of the square of the overlap of the deuteron
with the triton wave functions is equal to$^{21}$ $P_d(^3\mbox{H})=0.445$
that indicates inadequacy of the simplest cluster model (7) for description
of the triton, more sophisticated models are needed in such a case.

In our work, to evaluate the magnitude of the electric dipole
polarizability of the hypertriton, we use the function $\Psi_0$ in
the form (7) accounting for the above-mentioned peculiarity of the
hypertriton structure with a high probability of finding two
nucleons in the state of the deuteron. Substituting (7) into the
rigorous three-particle formula (5) for
$\alpha_E(^3_{\Lambda}\mbox{H})$, the expression (5) takes the form
\begin{eqnarray}
\alpha_E(^3_{\Lambda}\mbox{H})=\frac{4}{3}\frac{e^2}{\hbar^2}
\frac{m_pm_n}{m_{pn}}\int_{0}^{\infty}\frac{p^2dp}{2\pi^2}\left\{ \left(
\frac{m_{\Lambda}}{m_{pn\Lambda}}\right)^2
I(p)[\phi_{\Lambda}^{\prime}(p)]^2\right. \nonumber \\ [1mm]
+\left.\left( \frac{m_n}{m_{pn}}\right)^2 J(p)
[\phi_{\Lambda}(p)]^2 \right\}\;, \label{line2}
\end{eqnarray}
where the functions $I(p)$ and $J(p)$ are defined by
\begin{eqnarray}
I(p)=\int_{0}^{\infty}\frac{dk k^2}{2\pi^2} \frac{[\psi_0(k)]^2}
{{k^2}+[C(p)]^2}\;,\quad J(p)=\int_{0}^{\infty} \frac{dk k^2}{2\pi^2}
\frac{[\psi_0^\prime(k)]^2}{{k^2}+[C(p)]^2}\;, \label{line1}\\ [2mm]
{}[C(p)]^2 \equiv \frac{m_pm_n}{m_{pn}m_{\Lambda}} \left\{ \frac{m_{pn\Lambda}}
{m_{pn}} p^2 +\frac{m_{d\Lambda}}{m_d} \kappa_{\Lambda}^2 \right\}
+\kappa_d^2\;\;. \nonumber
\end{eqnarray}

The wave functions $\psi_0(k)$ and $\phi_{\Lambda}(p)$ in (8) and
(9) were determined by solving the corresponding two-body problems
with S-wave separable interaction potentials --- for the system
proton + neutron with the interaction potential $v_{pn}$ ,
\begin{equation}
v_{pn}(k,k^\prime)=-\gamma_{pn} w_{pn}(k) w_{pn}(k^\prime)
\end{equation}
and for the $\Lambda$ hyperon in a field of the effective
interaction potential $v_{\Lambda}$,
\begin{equation}
v_{\Lambda}(p,p^\prime)=-\gamma_{\Lambda} w_{\Lambda}(p) w_{\Lambda}(p^\prime),
\end{equation}
where the parameters $\gamma_{pn}$ and $\gamma_{\Lambda}$
characterize the strength of the corresponding interaction and the
formfactors $w_{pn}(k)$ and $w_{\Lambda}(p)$ describe dependences
of the interaction potentials on the variable momenta $k$ and $p$.
We use the $p-n$ and $\Lambda-d$ interaction potentials, (10) and (11),
with the formfactors of three different kinds --- Yukawa formfactor (Yu),
exponential (in configuration space) formfactor (Exp) and the formfactor
corresponding to the delta-shell potential (DS),
\begin{equation}
w^{\mbox {\scriptsize Yu}}(k)=\frac{\beta^2}{k^2+\beta^2},\;\;
w^{\mbox {\scriptsize Exp}}(k)=\frac{\beta^4}{(k^2+\beta^2)^2},\;\;
w^{\mbox {\scriptsize DS}}(k)=\frac{sin(k/\beta)}{(k/\beta)}.
\end{equation}
All the formfactors are normalized to one at zero momentum, $w(0)=1$.

With a certain formfactor, each of the potentials ($v_{pn}$ and
$v_{\Lambda}$) is given by two parameters which characterize the
strength and the inverse radius of interaction ($\gamma_{pn},
\beta_{pn}$ and $\gamma_{\Lambda}, \beta_{\Lambda}$). The model of
zero-range interaction (ZR) follows immediately from the model of
separable potential with any one of the formfactors (12) providing
the parameter of the inverse radius of interaction takes an infinite
large value: $\beta\rightarrow\infty$.\\

\vspace*{.1in}
\noindent {\bf 4. Calculations and discussion of results} \\ [.1in]
Calculations of the electric dipole polarizability of the
hypertriton $\alpha_E(^3_{\Lambda}\mbox{H})$ were carried out by the
formulae (8) and (9). The two-body wave functions $\psi_0(k)$ and
$\phi_{\Lambda}(p)$ were found using both the model of zero-range
interaction (ZR) and the model of the separable S-wave interaction
(10) and (11) with the different formfactors (12).

The parameters of the  $p-n$ interaction potentials were fitted to
the experimental values of the deuteron binding energy $B_d$ (Ref. 22)
and the triplet $p-n$ scattering length $^3a_{pn}$ (Ref. 23),
\begin{equation}
B_d=2.224575(9)\;\mbox{MeV}\;\; \mbox{(Ref. 22)}\,,\;\;
^3a_{pn}=5.424(3)\;\mbox{fm}\;\; \mbox{(Ref. 23)}.
\end{equation}

Of the parameters needed for fixing the low-energy effective
$\Lambda-d$ interaction potential (11) with the formfactors
(12), it is known to date only one --- the binding energy of the
$\Lambda$ hyperon in the hypertriton $B_{\Lambda}$,
\begin{equation}
B_{\Lambda}=0.13(5)\;\mbox{MeV}\;\; \mbox{(Ref. 24)}\;.
\end{equation}

Relying on just one parameter --- on the experimental value of the
binding energy $B_{\Lambda}$ --- the $\Lambda-d$ interaction
can be described only at very low energies by applying the zero range
interaction model.

To evaluate the influence of finiteness of the effective $\Lambda-d$
interaction radius on the magnitude of the polarizability
$\alpha_E(^3_\Lambda\mbox{H})$ it is worthy to utilize as an
second parameter the theoretical value of the doublet $\Lambda$
hyperon - deuteron scattering length $^2a_{{\Lambda}d}$ calculated
by a number of authors on the basis of the three-particle
description of the hypertriton as a bound system of the proton,
the neutron and the $\Lambda$ hyperon with the use of the
low-energy data for $p-n$ and $\Lambda$-nucleon interactions (see
Ref. 25 and citations therein). The results of analysis in Refs.
25 and 26 point to the existence of a simple correlation
dependence between $B_{\Lambda}$ and $^2a_{{\Lambda}d}$ that
follows readily from the effective range theory of the
$\Lambda-d$ interaction under condition that the $B_{\Lambda}$
is much smaller than the deuteron binding energy $B_d$. According
to the correlation, to the experimental valueÿ $B_{\Lambda}$ (14)
there corresponds$^{25}$ the doublet scattering length
\begin{equation}
^2a_{{\Lambda}d}=15.9\;\; \mbox{fm}\;.
\end{equation}
Notice that the value (15) is consistent with the values calculated
recently in the framework of the effective field theory in Ref. 27.

The fitted values of the parameters of the interaction potentials
$v_{pn}$ and $v_{\Lambda}$ ($\gamma_{pn}, \beta_{pn}$ and $\gamma_{\Lambda},
\beta_{\Lambda}$) are presented in the Table 1 for the various formfactors
$w_{pn}$ and $w_{\Lambda}$ (12). \\

{\footnotesize \tablename\hspace{2mm}1.\hspace{1mm} The parameters of the
potentials $v_{pn}$ and $v_\Lambda$, (8) and (9), with the formfactors
(12) that correspond to the experimental values of the deuteron binding
energy $B_d$ and the triplet $p-n$ scattering length $^3a_{np}$ (13)
(for $p-n$ interaction) and to the experimental binding energy of
the $\Lambda$ hyperon in the hypertriton $B_{\Lambda}$ (14) and the
theoretically evaluated value of the doublet $\Lambda-d$ scattering
length $^2a_{{\Lambda}d}=15.9\;\; \mbox{fm}^{25}$ (for $\Lambda-d$
interaction).
\begin{center} \begin{tabular}{|c|c|c|c|c|} \hline
\multicolumn{1}{|c|}{\hspace*{\fill}}&
\multicolumn{2}{c|}{System $p+n$}&
\multicolumn{2}{c|}{System $\Lambda+d$}\\ \cline{2-5}
\multicolumn{1}{|c}{}&
\multicolumn{1}{|c|}{}&
\multicolumn{1}{c|}{}&
\multicolumn{1}{c|}{}&
\multicolumn{1}{c|}{}\\
\multicolumn{1}{|c}{Formfactor}&
\multicolumn{1}{|c|}{$2\mu_{pn}\gamma_{pn},\mbox{fm}$}&
\multicolumn{1}{c|}{$\beta_{pn},\mbox{fm}^{-1}$}&
\multicolumn{1}{c|}{$2\mu_{{\Lambda}d}\gamma_{\Lambda},\mbox{fm}$}&
\multicolumn{1}{c|}{$\beta_{\Lambda},\mbox{fm}^{-1}$}\\
\multicolumn{1}{|c}{}&
\multicolumn{1}{|c|}{}&
\multicolumn{1}{c|}{}&
\multicolumn{1}{c|}{}&
\multicolumn{1}{c|}{}\\  \hline
{}&{}&{}&{}&\\
Yu&24.5955 &1.3906&23.5403&1.1934\\
Exp&27.5011&2.0522&25.9893&1.7485\\
DS&27.7483&0.6372&26.5333&0.5365\\ \hline
\end{tabular}
\end{center}}
\bigskip

In the case of using the zero-range model for both interactions
(between the proton and the neutron and between the $\Lambda$ hyperon
and the deuteron as a whole), based oneself upon two experimental
parameters --- the deuteron binding energy $B_d$ (13) and the $\Lambda$
hyperon binding energy in the hypertriton $B_\Lambda$ (14) --- we find
for the electric dipole  polarizability of the hypertriton the value
\begin{equation}
\alpha_E^{\mbox{{\scriptsize ZR,ZR}}}(^3_{\Lambda}\mbox{H})=1.701\;\;
\mbox{fm}^3\,.
\end{equation}

Taking into account the finite value of the proton-neutron interaction
radius through the application of the separable potential (8), whereas
still describing the effective interaction between the $\Lambda$ hyperon
and the deuteron with the help of the zero-range interaction model, we
obtain based oneself upon the three experimental parameters --- $B_d\;,
\;^3a_{pn}$ (11) and $B_{\Lambda}$ (12) --- a larger magnitude of the
hypertriton polarizability. For example, with the use of the Yukawa
formfactor for the $n-p$ interaction potential and the zero-range model
for the $\Lambda-d$ interaction, we find
\begin{equation}
\alpha_E^{\mbox{{\scriptsize Yu,ZR}}}(^3_{\Lambda}\mbox{H})=2.479\;\;
\mbox{fm}^3\,,
\end{equation}

To evaluate the effect of the finite-range character of the effective
$\Lambda-d$ interaction, it should be introduced one more parameter
in addition. Taking into the consideration that no low-energy parameter
characterizing the $\Lambda-d$ interaction other than $B_\Lambda$ is
experimentally determined, it is expedient to use the theoretical value of the
doublet $\Lambda-d$ scattering length (15) as an additional parameter.

Tabulated in the Tables 2 and 3 are the calculated values of the electric
dipole polarizabilities of the deuteron and the hypertriton,
$\alpha_E(^2\mbox{H})$ and $\alpha_E(^3_{\Lambda}\mbox{H})$ (in
$\mbox{fm}^3$), together with the geometrical characteristics of the
nuclei (in $\mbox{fm}$) obtained with the use of the zero-range model
and the separable potentials (10), (11) (with the formfactors (12) and the
parameters given in the Table 1) for the description of the
$p-n$ and ${\Lambda}-d$ interactions.\\

{\footnotesize \tablename\hspace{2mm}2.\hspace{1mm} The electric
dipole polarizability of the deuteron $\alpha_E(^2\mbox{H})$ (in
$\mbox{fm}^3$) and the root-mean-square distances
of the nucleons of the deuteron to its centre of mass,
$R_p$ and $R_n$, and between the nucleons of the deuteron,
$R_{pn}$, (in $\mbox{fm}$) calculated for various potentials.
\begin{center} \begin{tabular}{|c|c|c|c|c|} \hline
\multicolumn{1}{|c|}{\hspace*{\fill}}&
\multicolumn{4}{c|}{$p-n$ interaction}\\ \cline{2-5}
\multicolumn{1}{|c|}{}& \multicolumn{1}{c|}{}& \multicolumn{1}{c|}{}&
\multicolumn{1}{c|}{}& \multicolumn{1}{c|}{} \\ \multicolumn{1}{|c|}{Quantity}&
\multicolumn{1}{c|}{$\mbox{ZR}$}&
\multicolumn{1}{c|}{$\mbox{Yu}$}&
\multicolumn{1}{c|}{$\mbox{Exp}$}&
\multicolumn{1}{c|}{$\mbox{DS}$} \\
\multicolumn{1}{|c|}{}&
\multicolumn{1}{c|}{}&
\multicolumn{1}{c|}{}&
\multicolumn{1}{c|}{}&
\multicolumn{1}{c|}{} \\ \hline
$\alpha_E(^2\mbox{H})$&0.378&0.626&0.627&0.629 \\
$R_p$&1.528&1.938&1.943&1.956 \\
$R_n$&1.525&1.935&1.941&1.953 \\
$R_{pn}$&3.053&3.873&3.884&3.908 \\ \hline
\end{tabular}
\end{center}}
\bigskip

{\footnotesize \tablename\hspace{2mm}3.\hspace{1mm} The electric
dipole polarizability of the hypertriton $\alpha_E(^3_{\Lambda}\mbox{H})$
(in $\mbox{fm}^3$) and the root-mean-square distances
between each of the hadrons of the hypertriton ($p, n, \Lambda$) and the
centre of mass of the hypertriton, $R_p$, $R_n$ and $R_{\Lambda}$, and
between the hadrons, $R_{pn}$, $R_{{\Lambda}p}$ and $R_{{\Lambda}n}$,
(in $\mbox{fm}$) calculated for various potentials.
\begin{center} \begin{tabular}{|c|c|c|c|c|c|c|} \hline
\multicolumn{1}{|c|}{\hspace*{\fill}}&
\multicolumn{6}{c|}{$p-n, \Lambda-d$ interactions}\\ \cline{2-7}
\multicolumn{1}{|c|}{}&
\multicolumn{1}{c|}{}&
\multicolumn{1}{c|}{}&
\multicolumn{1}{c|}{}&
\multicolumn{1}{c|}{}&
\multicolumn{1}{c|}{}&
\multicolumn{1}{c|}{} \\
\multicolumn{1}{|c|}{Quantity}&
\multicolumn{1}{c|}{$\mbox{ZR,ZR}$}&
\multicolumn{1}{c|}{$\mbox{Yu,ZR}$}&
\multicolumn{1}{c|}{$\mbox{ZR,Yu}$}&
\multicolumn{1}{c|}{$\mbox{Yu,Yu}$}&
\multicolumn{1}{c|}{$\mbox{Exp,Exp}$}&
\multicolumn{1}{c|}{$\mbox{DS,DS}$} \\
\multicolumn{1}{|c|}{}&
\multicolumn{1}{c|}{}&
\multicolumn{1}{c|}{}&
\multicolumn{1}{c|}{}&
\multicolumn{1}{c|}{}&
\multicolumn{1}{c|}{}&
\multicolumn{1}{c|}{} \\ \hline
$\alpha_E(^3_{\Lambda}\mbox{H})$&1.701&2.479&2.002&2.914&2.933&2.964 \\
$R_p$&4.148&4.315&4.470&4.626&4.629&4.636 \\
$R_n$&4.147&4.314&4.469&4.625&4.628&4.635 \\
$R_{\Lambda}$&6.490&4.490&7.070&7.070&7.072&7.074 \\
$R_{pn}$&3.053&3.873&3.053&3.873&3.884&3.908 \\
$R_{{\Lambda}p}$&10.458&10.526&11.374&11.436&11.439&11.446 \\
$R_{{\Lambda}n}$&10.458&10.525&11.374&11.436&11.439&11.445 \\
\hline
\end{tabular}
\end{center}}
\bigskip

Note that in the zero-range interaction limit the hypertriton is
found to be a more compact system than with the use of more
realistic interaction which is characterized by a finite range.
This is also evident from the calculated geometrical
characteristics of the system given in the Table 3. The more
compact hypertriton system is certain to possess the lesser value
of the electric polarizability.

For comparison, the electric dipole polarizability of the
deuteron, calculated in the zero-range $p-n$ interaction limit,
$\alpha_E^{{\mbox{\scriptsize ZR}}}(^2\mbox{H})=0.378\;\mbox{fm}^3$,
is also decreased as against the values obtained
with allowance for the finite magnitude of the range of the
interaction, which are found to be  $\alpha_E^{\mbox{{\scriptsize
Yu}}}(^2\mbox{H})=0.626\;\; \mbox{fm}^3$ for the Yukawa
formfactor, $\alpha_E^{\mbox{{\scriptsize
Exp}}}(^2\mbox{H})=0.627\;\; \mbox{fm}^3$ for the exponential
formfactor and $\alpha_E^{\mbox{{\scriptsize DS}}}
(^2\mbox{H})=0.629\;\; \mbox{fm}^3$ for the delta-shell
formfactor.

It follows from the results of our calculations presented in the
Table 3 that taking into account the finite-range character of the
$p-n$ interaction has a markedly more effect on the magnitude of
the hypertriton electric polarizability
$\alpha_E(^3_{\Lambda}\mbox{H})$ than that of the $\Lambda-d$
interaction: $\alpha_E^{\mbox{{\scriptsize
Yu,ZR}}}(^3_{\Lambda}\mbox{H})>\alpha_E^{\mbox{{\scriptsize
ZR,Yu}}}(^3_{\Lambda}\mbox{H})$. This is due to the fact that the
leading term of the effective range expansion (zero-range
approximation) describes better the low-energy $\Lambda-d$
scattering as compared to the low-energy $p-n$ scattering
because of the different extent of proximity to bound state pole in the
corresponding transition matrices (the binding energy of the
$\Lambda$ hyperon in the hypertriton $B_{\Lambda}$ is noticeably
lesser than the deuteron binding energy $B_d$).

The calculated magnitude of the electric polarizability of the
hypertriton depends only slightly on the kind of the used
formfactor of the separable potential, varying from the value
$\alpha_E^{\mbox{{\scriptsize Yu,Yu}}}(^3_{\Lambda}\mbox{H})=2.914\;
\mbox{fm}^3$ for the Yukawa formfactor to the value
$\alpha_E^{\mbox{{\scriptsize DS,DS}}}(^3_{\Lambda}\mbox{H})=2.964\;
\mbox{fm}^3$ for the delta-shell one. Hence, from our results
obtained it may be inferred that the value of the hypertriton
polarizability $\alpha_E(^3_{\Lambda}\mbox{H})$ is close to $3\;
\mbox{fm}^3$ exceeding the polarizabilities of the ordinary
three-nucleon nuclei by an order of magnitude.

The Tables 2 and 3 list also geometrical characteristics of the
deuteron and hypertriton, calculated with the use of the deuteron
wave function $\psi_0(k)$ and the hypertriton wave function in the
form (7): the root-mean-square distances between the corresponding
particles and the centre of mass of the nuclei ($R_p$, $R_n$ and
$R_{\Lambda}$), together with the root-mean-square distances
between the proton and the neutron ($R_{pn}$), between the
$\Lambda$ hyperon and the proton ($R_{{\Lambda}p}$) and between
the $\Lambda$ hyperon and the neutron ($R_{{\Lambda}n}$). Note
that in the case of the application of the model wave function (7)
the distance between the proton and the neutron in the hypertriton
remains identical to that in the deuteron,
$R_{pn}(^3_{\Lambda}\mbox{H}) =R_{pn}(^2\mbox{H})$ , where
$R_{pn}(^2\mbox{H})=R_p(^2\mbox{H})+ R_n(^2\mbox{H})$. That there
is only inconsiderable deformation of the deuteron in the
hypertriton follows from the three-body calculations by Kolesnikov
and Kalachev $^{28}$:
$R_{pn}(^3_{\Lambda}\mbox{H})/R_{pn}(^2\mbox{H})\approx0.9$.
According to our calculations, the root-mean-square distance
between the $\Lambda$ hyperon and the centre of mass of the
hypertriton, $R_{\Lambda}$, is more than half as large again as
the corresponding distances of the proton and the neutron, $R_p$
and $R_n$. Hence, in the hypertriton the distances between the
$\Lambda$ hyperon and the proton and between the $\Lambda$ hyperon
and the neutron, $R_{{\Lambda}p}$ and $R_{{\Lambda}n}$, are more
than twice as large as between the proton and neutron, $R_{pn}$.\\

\vspace*{.1in}
\noindent {\bf 5. Summary and conclusions} \\[.1in]
On the basis of the few-body approach, a rigorous formalism for
determination of the electric dipole polarizability of a
three-hadron bound complex that consists of one charged and two
neutral particles and can form only one stable bound state has
been worked out. Leaning upon the analytical structure of the
three-body transition matrix, a simple expression for the electric
dipole polarizability of the hypertriton in terms of the partial
derivatives of its wave function in momentum space has been
formulated (Eq. (5)).

In the framework of the developed formalism, taking into
consideration the specific structure of the hypertriton as a
strange halo nucleus consisting of a tightly bound core (a
$^2\mbox{H}$ nucleus) surrounded by the loosely bound $\Lambda$ hyperon,
the electric dipole polarizability of the hypertriton
$\alpha_E(^3_{\Lambda}\mbox{H})$ has been firstly calculated.

With the use of the $p-n$ separable interaction potential with
the Yukawa formfactor and the $\Lambda-d$ zero-range interaction
corresponding to the three experimentally established low-energy quantities ---
the binding energy of the deuteron $B_d$, the triplet $p-n$ scattering length
$^3a_{pn}$ and the binding energy of the $\Lambda$ hyperon in the hypertriton
$B_{\Lambda}$ --- the value of the electric polarizability of the hypertriton
is found to be $\alpha_E^{\mbox{{\mbox{\scriptsize
Yu,ZR}}}}(^3_{\Lambda}\mbox{H}) =2.48\;\; \mbox{fm}^3$.

The evaluation of the finite-range effects of the $\Lambda-d$
interaction has been carried out using an additional theoretically
calculated low-energy quantity --- the doublet $\Lambda-d$ scattering
length $^2a_{{\Lambda}d}$.  It has been revealed that taking into
account the finite-range character of the $\Lambda-d$ interaction
causes a less pronounced increase of the hypertriton polarizability
$\alpha_E(^3_{\Lambda}\mbox{H}$) than that of the $p-n$ interaction.
With the use of all the four $p-n$ and $\Lambda-d$ low-energy parameters,
the calculated magnitude of the electric polarizability of the hypertriton
$\alpha_E(^3_{\Lambda}\mbox{H}$) depends only slightly on the kind of the
formfactor of the $p-n$ and $\Lambda-d$ interaction potentials taking values
close to $3\;\;\mbox{fm}^3$ (Table 3).

Thus it follows from our study that the electric polarizability of the
$\Lambda$ hypertriton $^3_{\Lambda}\mbox{H}$ by an order of magnitude
exceeds the the polarizabilities of the ordinary three-nucleon systems ---
the triton and the helion-3. A further refinement of the hypertriton
polarizability can be implemented with the use of three-body calculational
techniques for obtaining the wave function of the bound state.\\

\vspace*{.1in}
\noindent {\footnotesize {\bf References}
\vspace*{.1in}
\begin{itemize}
\setlength{\baselineskip}{.1in}
\item[{\tt 1.}]N. L. Rodning, L. D.Knutson, W. G. Lynch, and M.
           B. Tsang,{\it Phys. Rev. Lett.} {\bf 49}, 909 (1982).
\item[{\tt 2.}]F. Goeckner, L. O. Lamm and L. D. Knutson,
           {\it Phys. Rev.} {\bf C43}, 66 (1991).
\item[{\tt 3.}]J. L. Friar, S. Fallieros, E. L. Tomusiak, D. Skopik and
           E. G. Fuller, {\it Phys. Rev.} {\bf C27}, 1364 (1983).
\item[{\tt 4.}]J. L. Friar and G. L. Payne, {\it Phys. Rev.} {\bf C55}, 2764
           (1997).
\item[{\tt 5.}]A. V. Kharchenko, {\it Nucl. Phys.} {\bf A617}, 34 (1997).
\item[{\tt 6.}]M. H. Lopes, J. A. Tostevin and R. C. Johnson, {\it Phys. Rev.}
           {\bf C28}, 1779 (1983).
\item[{\tt 7.}]O. G. Sitenko and A. V. Kharchenko, {\it Ukrainian J. Phys.}
           {\bf 42}, 798 (1997).
\item[{\tt 8.}]J.-W. Chen, H. W. Grie{\ss}hammer, M. J. Savage and
           R. P. Springer, {\it Nucl. Phys.} {\bf A644}, 221 (1998);
           arXiv:nucl-th/9806080.
\item[{\tt 9.}]X. Ji and Y. Li, {\it Phys. Lett.} {\bf B591}, 76 (2004).
\item[{\tt 10.}]J. K. Ahn {\it et al.}, {\it Phys. Rev. Lett.} {\bf 87},
           132504 (2001).
\item[{\tt 11.}]D. J. Prowse {\it et al.}, {\it Phys. Rev. Lett.} {\bf 17},
           782 (1966).
\item[{\tt 12.}]H. Takahashi {\it et al.}, {\it Phys. Rev. Lett.} {\bf 87},
           212502 (2001).
\item[{\tt 13.}]S. D. Randenia and E. V. Hungeford, {\it Phys. Rev.}
           {\bf C76}, 064308 (2007).
\item[{\tt 14.}]V. F. Kharchenko and S. A. Shadchin, {\it Few-Body Systems}
           {\bf 6}, 45 (1989); {\it Yad. Fiz.} {\bf 45}, 333 (1987).
\item[{\tt 15.}]V. F. Kharchenko and S. A. Shadchin S.A., {\it Three-body
           theory of the effective interaction between a particle and a two-
           particle bound system}, preprint ITP-93-24E (Institute for
           Theoretical Physics, Kyiv, 1993).
\item[{\tt 16.}]V. F. Kharchenko and S. A. Shadchin, {\it Ukrainian J. Phys.}
           {\bf 42}, 11 (1997).
\item[{\tt 17.}]V. F. Kharchenko, {\it J. Phys. Studies} {\bf 4},
           245 (2000).
\item[{\tt 18.}]V. F. Kharchenko, S. A. Shadchin and
S. A. Permyakov, {\it
           Phys. Lett.} {\bf B199}, 1 (1987).
\item[{\tt 19.}]V. F. Kharchenko and S. A. Shadchin, {\it Ukrainian J. Phys.}
           {\bf 42}, 912 (1997).
\item[{\tt 20.}]L. D. Faddeev, {\it Zh. Eksp. Teor. Fiz.} {\bf 39}, 1459 (1960).
\item[{\tt 21.}]K. Miyagawa, H. Kamada, W. Gl\"{o}ckle and V. Stoks, {\it Phys.
           Rev.} {\bf C51}, 2905 (1995).
\item[{\tt 22.}]C. Van der Leun and C. Alderliesten, {\it Nucl. Phys.}
           {\bf A380}, 261 (1982).
\item[{\tt 23.}]L. Koester and W. Nistler, {\it Z. Phys.} {\bf A272}, 189
           (1975).
\item[{\tt 24.}]M. Juri\'{c} {\it et al.}, {\it Nucl. Phys.} {\bf B52}, 1
           (1973).
\item[{\tt 25.}]V. V. Peresypkin and N. M. Petrov, {\it Binding energy of
           hypertriton and doublet scattering length of
           $\Lambda$-hyperon-deuteron scattering for nonlocal separable
           potentials}, preprint ITF-75-39R (Institute for Theoretical Physics,
           Kyiv, 1975).
\item[{\tt 26.}]N. M. Petrov, {\it Yad. Fiz.} {\bf 48}, 50 (1988).
\item[{\tt 27.}]H. Garcilazo, T. Fern\'{a}ndez-Caram\'{e}s and A. Valcarce,
           {\it Phys. Rev} {\bf C75}, 034002 (2007).
\item[{\tt 28.}]N. N. Kolesnikov and S. A. Kalachev, {\it Yad. Fiz.} {\bf 69},
           2064 (2006).

\end{itemize}}

\end{document}